\documentclass[conference]{IEEEtran}

\usepackage{graphicx}
\usepackage{subcaption}
\usepackage{tikz}
\usepackage[nolist]{acronym} 
\usepackage{pgfplots}
\pgfplotsset{compat=newest}
\usepackage[bookmarks=false]{hyperref}
\usepackage{units}
\usepackage{amsmath, amsbsy, amssymb}

\usepackage{xcolor}
\definecolor{mittelblau}{RGB}{0, 126, 198}
\definecolor{violettblau}{cmyk}{0.9, 0.6, 0, 0}
\definecolor{rot}{RGB}{238, 28 35}
\definecolor{apfelgruen}{RGB}{140, 198, 62}
\definecolor{gelb}{RGB}{255, 229, 0}
\definecolor{orange}{RGB}{244, 111, 33}
\definecolor{pink}{RGB}{237, 0, 140}
\definecolor{lila}{RGB}{128, 10, 145}
\definecolor{hellgrau}{RGB}{224, 224, 224}
\definecolor{mittelgrau}{RGB}{128, 128, 128}
\definecolor{dunkelgrau}{RGB}{80,80,80}
\definecolor{anthrazit}{RGB}{19, 31, 31}

\begin{document}

\title{On Deep Learning-Based Channel Decoding}

\author{\IEEEauthorblockN{Tobias Gruber\IEEEauthorrefmark{1}, Sebastian Cammerer\IEEEauthorrefmark{1}, Jakob Hoydis\IEEEauthorrefmark{2}, and Stephan ten Brink\IEEEauthorrefmark{1} }
\IEEEauthorblockA{
\IEEEauthorrefmark{1} Institute of Telecommunications, Pfaffenwaldring 47, University of  Stuttgart, 70659 Stuttgart, Germany 
    \\\{gruber,cammerer,tenbrink\}@inue.uni-stuttgart.de
\\
\IEEEauthorrefmark{2}Nokia Bell Labs, Route de Villejust, 91620 Nozay, France
    \\jakob.hoydis@nokia-bell-labs.com
}
}

\maketitle

\begin{acronym}
 \acro{NN}{neural network}
 \acro{ECC}{error-correcting code}
 \acro{MLD}{maximum likelihood decoding}
 \acro{HDD}{hard decision decoding}
 \acro{SDD}{soft decision decoding}
 \acro{NND}{neural network decoding}
 \acro{ML}{maximum likelihood}
 \acro{GPU}{graphical processing unit}
 \acro{BP}{belief propagation}
 \acro{LDPC}{low density parity check}
 \acro{BER}{bit error rate}
 \acro{SNR}{signal-to-noise-ratio}
 \acro{ReLU}{rectified linear unit}
 \acro{BPSK}{binary phase shift keying}
 \acro{AWGN}{additive white Gaussian noise}
 \acro{MSE}{mean squared error}
 \acro{LLR}{log-likelihood ratio}
 \acro{MAP}{maximum a posteriori}
 \acro{NVE}{normalized validation error}
 \acro{BCE}{binary cross-entropy}
 \acro{BLER}{block error rate}
 \acro{IoT}{internet of things}
\end{acronym}

\begin{abstract}
 We revisit the idea of using deep neural networks for one-shot decoding of random and structured codes, such as polar codes. Although it is possible to achieve \ac{MAP} \ac{BER} performance for both code families and for short codeword lengths, we observe that (i) structured codes are easier to learn and (ii) the neural network is able to generalize to codewords that it has never seen during training for structured, but not for random codes. These results provide some evidence that neural networks can learn a form of decoding algorithm, rather than only a simple classifier. We introduce the metric \emph{\ac{NVE}} in order to further investigate the potential and limitations of deep learning-based decoding with respect to performance and complexity.
\end{abstract}

\acresetall
\section{Introduction}

\begin{figure*}[t]
 \centering
 \resizebox{\textwidth}{!}{
\resizebox{\linewidth}{!}{
\begin{tikzpicture}
	\tikzstyle{hidden} = [draw, circle, inner sep=3pt, fill=gray] 
    \tikzstyle{input} = [draw, circle, inner sep=3pt, fill=black]
    \tikzstyle{output} = [circle, inner sep=3pt, fill=white]
    \tikzstyle{dots}=[circle, inner sep=0.5pt, fill=black]

  \foreach \i in {1,2} {
		\coordinate (srco-\i) at (-10,0.3+0.5*\i);
		\coordinate (srcu-\i) at (-10,-0.3-0.5*\i);
	}

    \foreach \i in {1,2,3} {
		\node[dots] at (-8.625,0.5-0.25*\i) {};
	}

   \foreach \i in {1,2,3} {
		\coordinate (enco-\i) at (-6.25,0.3+0.5*\i);
		\coordinate (encu-\i) at (-6.25,-0.3-0.5*\i);
	}

\foreach \i in {1,2,3} {
		\node[dots] at (-3.75,0.5-0.25*\i) {};
	}

	\foreach \i in {1,2} {
		\draw[-] (srco-\i) -- (enco-\i);
		\draw[-] (srcu-\i) -- (encu-\i); 
	}


\draw[fill=white] (-10,0) ellipse (0.5cm and 3.25cm);
	\node[rotate=90] at (-10,0) {\large $k$ information bits};

	\foreach \i in {1,2,3} {
		\coordinate (starto-\i) at (-2.5,0.3+0.5*\i);
		\coordinate (startu-\i) at (-2.5,-0.3-0.5*\i);
	}

	\foreach \i in {1,2,3} {
		\node[dots] at (-1.25,0.5-0.25*\i) {};
	}

  	\foreach \i in {1,2,3} {
		\draw[-] (enco-\i) -- (starto-\i);
		\draw[-] (encu-\i) -- (startu-\i); 
	}	 
  
    \foreach \i in {1,2,3} {
		\node[input] (ino-\i) at (0,0.3+0.5*\i) {};
		\node[input] (inu-\i) at (0,-0.3-0.5*\i) {};
    }

	\foreach \i in {1,2,3} {
		\node[dots] at (0,0.5-0.25*\i) {};
	}

	\foreach \i in {1,2,3} {
		\draw[-] (starto-\i) -- (ino-\i);
		\draw[-] (startu-\i) -- (inu-\i); 
	}

	\node[draw, minimum width=2.2cm, text centered, minimum height=0.6cm,thick] at (0,4.3) {Input};
    
    \foreach \i in {1,2,3} {
    	\node[input] (modo-\i) at (1.5,0.3+0.5*\i) {};    
		\node[input] (modu-\i) at (1.5,-0.3-0.5*\i) {};  
    }

	\foreach \i in {1,2,3} {
		\node[dots] at (1.5,0.5-0.25*\i) {};
	}

	\foreach \i in {1,2,3} {
		\draw[-] (ino-\i) -- (modo-\i);
		\draw[-] (inu-\i) -- (modu-\i); 
	}

	\node[draw, minimum width=2.2cm, text centered, minimum height=0.6cm,color=rot,thick] at (1.5,3.6) {Modulation};
    
	\foreach \i in {1,2,3} {
    	\node[input] (no-\i) at (3,0.3+0.5*\i) {};
		\node[input] (nu-\i) at (3,-0.3-0.5*\i) {}; 
    }

	\foreach \i in {1,2,3} {
		\node[dots] at (3,0.5-0.25*\i) {};
	}

	\foreach \i in {1,2,3} {
		\draw[-] (modo-\i) -- (no-\i);
		\draw[-] (modu-\i) -- (nu-\i); 
	}

	\node[draw, minimum width=2.2cm, text centered, minimum height=0.6cm,color=rot,thick] at (3,4.3) {Noise};

	\foreach \i in {1,2,3} {
    	\node[input,fill=white, densely dashed] (llro-\i) at (4.5,0.3+0.5*\i) {};
		\node[input,fill=white,densely dashed] (llru-\i) at (4.5,-0.3-0.5*\i) {}; 
    }

	\foreach \i in {1,2,3} {
		\node[dots] at (4.5,0.5-0.25*\i) {};
	}

	\foreach \i in {1,2,3} {
		\draw[-] (no-\i) -- (llro-\i);
		\draw[-] (nu-\i) -- (llru-\i); 
	}

	\node[draw, minimum width=2.2cm, text centered, minimum height=0.6cm,color=rot, densely dashed,thick] at (4.5,3.6) {[optional] LLR};

	\foreach \i in {1,2,3} {
    	\node[input] (nndo-\i) at (6,0.3+0.5*\i) {};
		\node[input] (nndu-\i) at (6,-0.3-0.5*\i) {}; 
    }

	\foreach \i in {1,2,3} {
		\node[dots] at (6.5,0.5-0.25*\i) {};
	}

	\foreach \i in {1,2,3} {
		\draw[-] (llro-\i) -- (nndo-\i);
		\draw[-] (llru-\i) -- (nndu-\i); 
	}

	\node[draw, minimum width=2.2cm, text centered, minimum height=0.6cm,color=mittelblau,thick] at (6,4.3) {NND input};

	\foreach \i in {1,...,5} {
    	\node[input] (hlo1-\i) at (7.5,0.3+0.5*\i) {};
		\node[input] (hlu1-\i) at (7.5,-0.3-0.5*\i) {};
    }

	\foreach \i in {1,2,3} {
		\node[dots] at (7.5,0.5-0.25*\i) {};
	}
	
	\foreach \i in {1,2,3} {
		\foreach \j in {1,...,5} {
			\draw[-] (nndo-\i) -- (hlo1-\j); 
			\draw[-] (nndu-\i) -- (hlu1-\j); 
			\draw[-] (nndo-\i) -- (hlu1-\j); 
			\draw[-] (nndu-\i) -- (hlo1-\j); 
			}
	}

	\node[draw, minimum width=2.2cm, text centered, minimum height=0.6cm,color=mittelblau,thick] at (7.5,3.6) {Hidden 1};

	\foreach \i in {1,...,5} {
    	\node[input] (hlo2-\i) at (9,0.3+0.5*\i) {};
		\node[input] (hlu2-\i) at (9,-0.3-0.5*\i) {};
    }

	\foreach \i in {1,2,3} {
		\node[dots] at (9,0.5-0.25*\i) {};
	}

	\foreach \i in {1,...,5} {
		\foreach \j in {1,...,5} {
			\draw[-] (hlo1-\i) -- (hlo2-\j); 
			\draw[-] (hlu1-\i) -- (hlu2-\j); 
			\draw[-] (hlo1-\i) -- (hlu2-\j); 
			\draw[-] (hlu1-\i) -- (hlo2-\j); 
			}
	}

	\node[draw, minimum width=2.2cm, text centered, minimum height=0.6cm,color=mittelblau,thick] at (9,4.3) {Hidden 2};
    
	\foreach \i in {1,...,4} {
    	\node[input] (hlo3-\i) at (10.5,0.3+0.5*\i) {};
		\node[input] (hlu3-\i) at (10.5,-0.3-0.5*\i) {};
    }

	\foreach \i in {1,2,3} {
		\node[dots] at (10.5,0.5-0.25*\i) {};
	}

	\foreach \i in {1,...,5} {
		\foreach \j in {1,...,4} {
			\draw[-] (hlo2-\i) -- (hlo3-\j); 
			\draw[-] (hlu2-\i) -- (hlu3-\j); 
			\draw[-] (hlo2-\i) -- (hlu3-\j); 
			\draw[-] (hlu2-\i) -- (hlo3-\j); 
			}
	}

	\node[draw, minimum width=2.2cm, text centered, minimum height=0.6cm,color=mittelblau,thick] at (10.5,3.6) {Hidden 3};

	\foreach \i in {1,...,2} {
    	\node[input] (outo-\i) at (12,0.3+0.5*\i) {};
		\node[input] (outu-\i) at (12,-0.3-0.5*\i) {};
    }

	\foreach \i in {1,2,3} {
		\node[dots] at (12,0.5-0.25*\i) {};
	}

	\foreach \i in {1,...,4} {
		\foreach \j in {1,2} {
			\draw[-] (hlo3-\i) -- (outo-\j); 
			\draw[-] (hlu3-\i) -- (outu-\j); 
			\draw[-] (hlo3-\i) -- (outu-\j); 
			\draw[-] (hlu3-\i) -- (outo-\j); 
			}
	}

	\foreach \i in {1,2} {
		\coordinate (endo-\i) at (14.5,0.3+0.5*\i);
		\draw[-] (outo-\i) -- (endo-\i);
		\coordinate (endu-\i) at (14.5,-0.3-0.5*\i);
		\draw[-] (outu-\i) -- (endu-\i);
	}

	\foreach \i in {1,2,3} {
		\node[dots] at (13.25,0.5-0.25*\i) {};
	}

	\node[draw, minimum width=2.2cm, text centered, minimum height=0.6cm,color=mittelblau,thick] at (12,4.3) {Output};

	\draw[color=rot,thick] (1,3.05) rectangle (5,-2.5);
	\node[anchor=north east,color=rot] at (4.5,3.05) {\large abstract channel};

	\draw[color=mittelblau,thick] (5.5,3.05) rectangle (12.5,-3.05);
	\node[anchor=north east,color=mittelblau, align=right] at (12.5,3.05) {\large NN decoder};

	\draw[color=apfelgruen,thick] (-0.625,3.175) rectangle (12.625,-3.175);
	\node[anchor=south west,color=apfelgruen] at (-0.5,-3.175) {\large training neural network};

	\draw[fill=white] (-2.25,0) ellipse (0.5cm and 3.25cm);
	\draw[fill=white] (14.5,0) ellipse (0.5cm and 3.25cm);

	\node[rotate=90] at (-2.25,0) {\large  codeword $\mathbf{x}_i \in \mathcal{X}$ of length $N$};
	\node[rotate=90] at (14.5,0) {\large $k$ estimated information bits};

   \node[anchor=south] at (-8.625,1.2) {$b_0$};
  \node[anchor=south] at (-8.625,0.7) {$b_1$};
  \node[anchor=south] at (-8.625,-0.9) {$b_{k-2}$};
  \node[anchor=south] at (-8.625,-1.4) {$b_{k-1}$};

  \node[anchor=south] at (13.25,1.2) {$\hat{b}_0$};
  \node[anchor=south] at (13.25,0.7) {$\hat{b}_1$};
  \node[anchor=south] at (13.25,-0.9) {$\hat{b}_{k-2}$};
  \node[anchor=south] at (13.25,-1.4) {$\hat{b}_{k-1}$};
  
  \node[anchor=south] at (-3.75,1.7) {$x_{i,0}$};
  \node[anchor=south] at (-3.75,1.2) {$x_{i,1}$};
  \node[anchor=south] at (-3.75,0.7) {$x_{i,2}$};
  \node[anchor=south] at (-3.75,-0.9) {$x_{i,N-3}$};
  \node[anchor=south] at (-3.75,-1.4) {$x_{i,N-2}$};
  \node[anchor=south] at (-3.75,-1.9) {$x_{i,N-1}$};

\node[draw,text centered, text width=3cm,minimum height=5cm, fill=white] at (-6.25,0) {Encoder $\left\{0,1\right\}^k \rightarrow \left\{0,1\right\}^N$};



     
 \end{tikzpicture}
}
 }
 \caption{Deep learning setup for channel coding.}
 \label{fig:nn_setup} 
 \vspace{-0.3cm}
\end{figure*}
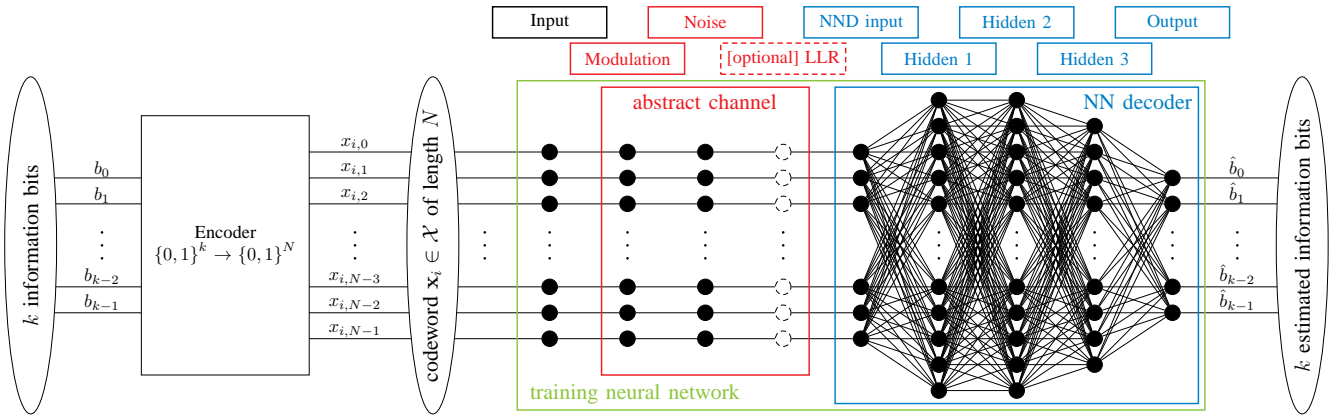

Deep learning-based channel decoding is doomed by the \emph{curse of dimensionality} \cite{wang1996}: for a short code of length ${N=100}$ and rate $r=0.5$, $2^{50}$ different codewords exist, which are far too many to fully train any \ac{NN} in practice. The only way that a \ac{NN} can be trained for practical blocklengths is, if it learns some form of decoding algorithm which can infer the full codebook from training on a small fraction of codewords. However, to be able to learn a decoding algorithm, the code itself must have some structure which is based on a simple encoding rule, like in the case of convolutional or algebraic codes. The goal of this paper is to shed some light on the question whether structured codes are easier to ``learn'' than random codes, and whether a \ac{NN} can decode codewords that it has never seen during training.

We want to emphasize that this work is based on very short blocklengths, i.e., $N\le 64$, which enables the comparison with \ac{MAP} decoding, but also has an independent interest for practical applications such as the \ac{IoT}. We are currently restricted to short codes because of the exponential training complexity \cite{wang1996}. Thus, the \ac{NND} concept is currently not competitive with state-of-the-art decoding algorithms which have been highly optimized over the last decades and scale to arbitrary blocklengths. 

Yet, there may be certain code structures which facilitate the learning process. One of our key finding is that structured codes are indeed easier to learn than random codes, i.e., less training epochs are required. Additionally, our results indicate that \acp{NN} may generalize or ``interpolate'' to the full codebook after having seen only a subset of examples, whenever the code has structure. 

\subsection{Related Work}

In 1943, McCulloch and Pitts published the idea of a \ac{NN} that models the architecture of the human brain in order to solve problems \cite{mcculloch1943}. But it took about 45 years until the backpropagation algorithm \cite{rumelhart1986} made useful applications such as handwritten ZIP code recognition possible \cite{lecun1989}. One early form of a \ac{NN} is a Hopfield net \cite{hopfield1982}. This concept was shown to be similar to \ac{MLD} of linear block \acp{ECC} \cite{bruck1989}: an erroneous codeword will converge to the nearest stable state of the Hopfield net which represents the most likely codeword. 
A naive implementation of \ac{MLD} means correlating the received vector of modulated symbols with all possible codewords which makes it infeasible for most practible codeword lengths, as the decoding complexity is $\mathcal{O}\left( 2^k \right)$ with $k$ denoting the number of information bits in the codeword. The parallel computing capabilities of \acp{NN} allow us to solve or, at least, approximate the \ac{MLD} problem in polynomial time \cite{zeng1989}. Moreover, the weights of the \ac{NN} are precomputed during training and the decoding step itself is then relatively simple. 


Due to its low storage capacity, Hopfield nets were soon replaced by feed-forward \acp{NN} which can learn an appropriate mapping between noisy input patterns and codewords. No assumption has to be made about the statistics of the channel noise because the \ac{NN} is able to learn the mapping or to extract the channel statistics during the learning process \cite{caid1990}. Different ideas around the use of \ac{NN} for decoding emerged in the 90s. While in \cite{caid1990} the output nodes represent the bits of the codeword, it is also possible to use one output node per codeword (\emph{one-hot} coding) \cite{stefano1991}. For Hamming coding, another variation is to use only the syndrome as input of the \ac{NN} in order to find the most likely error pattern \cite{tallini1995}. Subsequently, \ac{NND} for convolutional codes arose in 1996 when Wang and Wicker showed that \ac{NND} matches the performance of an ideal Viterbi decoder \cite{wang1996}. But they also mentioned a very important drawback of \ac{NND}: decoding problems have far more possibilities than conventional pattern recognition problems. This limits the \ac{NND} to short codes. However, the \ac{NN} decoder for convolutional codes was further improved by using recurrent neural nets \cite{hamalainen1999}.

\ac{NND} did not achieve any big breakthrough for neither block nor convolutional codes. Due to the standard training techniques in those times it was not possible to work with \acp{NN} employing a large number of neurons and layers, which rendered them unsuited for longer codewords. Hence, the interest in \acp{NN} dwindled, not only for machine learning applications but also for decoding purposes. Some slight improvements were made in the following years, e.g., by using random neural nets \cite{abdelbaki1999} or by reducing the number of weights \cite{wu2002}.

In 2006, a new training technique, called layer-by-layer unsupervised pre-training followed by gradient descent fine-tuning \cite{hinton2006}, led to the renaissance of \acp{NN} because it made training of \acp{NN} with more layers feasible. \acp{NN} with many hidden layers are called \emph{deep}. Nowadays, powerful new hardware such as \acp{GPU} are available to speed up learning as well as inference. In this renaissance of \acp{NN}, new \ac{NND} ideas emerge. Yet, compared to previous work, the \ac{NN} learning techniques are only used to optimize well known decoding schemes which we denote as introduction of expert knowledge. For instance, in \cite{nachmani2016}, weights are assigned to the Tanner graph of the \ac{BP} algorithm and learned by \ac{NN} techniques in order to improve the \ac{BP} algorithm. It still seems that the recent advances in the machine learning community have not yet been adapted to the pure idea of \emph{learning to decode}.

\section{Deep Learning for Channel Coding}
The theory of deep learning is comprehensively described in \cite{goodfellow2016}. Nevertheless, for completeness, we will briefly explain the main ideas and concepts in order to introduce a \ac{NN} for channel (de-)coding and its terminology. A \ac{NN} consists of many connected neurons. In such a neuron all of its weighted inputs are added up, a bias is optionally added, and the result is propagated through a nonlinear activation function, e.g., a sigmoid function or a \ac{ReLU}, which are respectively defined as
\begin{align}
  \operatorname{g}_{\text{sigmoid}} \left(z\right) &= \frac{1}{1+e^{-z}} , \qquad   \operatorname{g}_{\text{relu}} \left(z\right) &= \max\left\{ 0,z \right\}.
\end{align}
If the neurons are arranged in layers without feedback connections we speak of a feedforward \ac{NN} because information flows through the net from the left to the right  without feedback (see Fig.~\ref{fig:nn_setup}). Each layer $i$ with $n_i$ inputs and $m_i$ outputs performs the mapping $\mathbf{f}^{\left(i \right)}: \mathbb{R}^{n_i} \rightarrow \mathbb{R}^{m_i}$ with the weights and biases of the neurons as parameters. Denoting $\boldsymbol{v}$ as input and $\boldsymbol{w}$ as output of the \ac{NN}, an input-output mapping is defined by a chain of functions depending on the set of parameters $\boldsymbol{\theta}$ by
\begin{align}\label{eq:mapping_function}  
\boldsymbol{w} =  \mathbf{f}\left( \boldsymbol{v}; \boldsymbol{\theta} \right)  = \mathbf{f}^{\left(L-1 \right)} \left( \mathbf{f}^{\left(L-2 \right)} \left( \ldots \left( \mathbf{f}^{\left(0 \right)} \left( \boldsymbol{v} \right)  \right) \right) \right)                                                                                                                                                                                                                                                                                                                                                                                                                                                                                                                                                                                                                                                                                                                                                                                                                                                                                                                                                                                                                                                                                                                                                                                                                                                                                                                                                                                                                                                                                                                                                         \end{align}where $L$ gives the number of layers and is also called \emph{depth}. It was shown in \cite{hornik1989} that such a multilayer \ac{NN} with $L=2$ and nonlinear activation functions can theoretically approximate any continuous function on a bounded region arbitrarily closely---if the number of neurons is large enough. 

In order to find the optimal weights of the \ac{NN}, a training set of known input-output mappings is required and a specific loss function has to be defined. By the use of gradient descent optimization methods and the backpropagation algorithm \cite{rumelhart1986}, weights of the \ac{NN} can be found which minimize the loss function over the training set. The goal of training is to enable the \ac{NN} to find the correct outputs for unseen inputs. This is called generalization. In order to quantify the generalization ability, the loss can be determined for a data set that has not been used for training, the so-called validation set.

In this work, we want to use a \ac{NN} for decoding of noisy codewords. At the transmitter, $k$ information bits are encoded into a codeword of length $N$. The coded bits are modulated and transmitted over a noisy channel. At the receiver, a noisy version of the codeword is received and the task of the decoder is to recover the corresponding information bits. In comparison to iterative decoding, the \ac{NN} finds its estimate by passing each layer only once. As this principle enables low-latency implementations, we term it \emph{one-shot} decoding.

Obtaining labeled training data is usually a very hard and expensive task for the field of machine learning. But using \ac{NN} for channel coding is special because we deal with man-made signals. Therefore, we are able to generate as many training samples as we like. Moreover, the desired \ac{NN} output, also denoted as label, is obtained for free because if noisy codewords are generated, the transmitted information bits are obviously known. For the sake of simplicity, \ac{BPSK} modulation and an \ac{AWGN} channel is used. Other channels can be adopted straightforwardly, and it is this flexibility that may be a particular advantage of \ac{NN}-based decoding.

In order to keep the training set small it is possible to extend the \ac{NN} with additional layers for modulating and adding noise (see Fig. \ref{fig:nn_setup}). These additional layers have no trainable parameters, i.e., they perform a certain action such as adding noise and propagate this value only to the node of the next layer with the same index. Instead of creating, and thus storing, many noisy versions of the same codeword, working on the noiseless codeword is sufficient. Thus, the training set $\mathcal{X}$ consists of all possible codewords $\mathbf{x}_i\in\mathbb{F}_2^{N}$ with $\mathbb{F}_2\in\{0,1\}$ (the labels being the corresponding information bits) and is given by $\mathcal{X} = \left\{ \mathbf{x}_0, \ldots, \mathbf{x}_{2^{k-1}} \right\}.$

As recommended in \cite{goodfellow2016}, each hidden layer employs a \ac{ReLU} activation function because it is nonlinear and at the same time very close to linear which helps during optimization. Since the output layer represents the information bits, a sigmoid function forces the output neurons to be in between zero and one, which can be interpreted as the probability that a ``1'' was transmitted. If the probability is close to the bit of the label, the loss should be incremented only slightly whereas large errors should result in a very large loss. Examples for such loss functions are the \ac{MSE} and the \ac{BCE}, defined respectively as
\begin{align}
 \text{L}_{\text{\ac{MSE}}} &= \frac{1}{k} \sum_i \left( b_i - \hat{b}_i \right)^2 \\  
 \text{L}_{\text{\ac{BCE}}} &= -\frac{1}{k} \sum_i \left[ b_i \ln\left(\hat{b}_i \right) + \left( 1 - b_i \right) \ln \left( 1 - \hat{b}_i \right)  \right] 
\end{align}
where $b_i \in \left\{ 0,1 \right\}$ is the $i$th target information bit (label) and $\hat{b}_i \in \left[ 0,1\right]$ the \ac{NN} soft estimate.  

There are some alternatives for this setup. First, \ac{LLR} values could be used instead of channel values. For \ac{BPSK} modulation over an \ac{AWGN} channel, these are obtained by \begin{align}
\text{LLR}\left(y\right) = \ln\frac{P\left( x=0 \vert y \right)}{P\left( x=1 \vert y \right)}  = \frac{2}{\sigma^2} y                                                                                                                                                             \end{align}
where $\sigma^2$ is the noise power and $y$ the received channel value. This processing step can be also implemented as an additional layer without any trainable parameters. Note,  that the noise variance must be known in this case and provided as an additional input to the \ac{NN}.\footnote{Inspired by the idea of spatial transformer networks \cite{jaderberg2015spatial}, one could alternatively use a second \ac{NN} to estimate $\sigma^2$ from the input and provide this estimate as an additional parameter to the \ac{LLR} layer.} Representing the information bits in the output layer as a \emph{one-hot}-coded vector of length $2^k$ is another variant. However, we refrain from this idea since it does not scale to large values of $k$. 
Freely available open-source machine learning libraries, such as Theano\footnote{\url{https://github.com/Theano/Theano}}, help to implement and train complex \ac{NN} models on fast concurrent \ac{GPU} architectures. We use Keras\footnote{\url{https://github.com/fchollet/keras}}  as a convenient high-level abstraction front-end for Theano. It allows to quickly deploy \acp{NN} from a very abstract point of view in the Python programming language that hides away a lot of the underlying complexity. As we support reproducible research, we have made parts of the source code of this paper available.\footnote{\url{https://github.com/gruberto/DL-ChannelDecoding}}

\section{Learn to Decode} \label{sec:learn_to_decode}
In the sequel, we will consider two different code families: random codes and structured codes, namely polar codes \cite{ArikanPolar09}. Both have codeword length $N=16$ and code rate $r=0.5$. While random codes are generated by randomly picking codewords from the codeword space with a Hamming distance larger than two, the generator matrix of polar codes of block size $N = 2^n$ is given by \begin{align}                                                                                                                                                                                                                                                                                                                                                               \mathbf{G}_N  = \mathbf{F}^{\otimes n}, \qquad \mathbf{F} = \left[ \begin{array}{ll} 1 & 0 \\ 1 & 1                                                                                                                                                                                                                                                                                                                                                                                                                                                                                                                                                                                                                                                                                                        \end{array}\right]                                                                                                                                                                                                                                                                                                                                    \end{align}
where $\mathbf{F}^{\otimes n}$ denotes the $n$th Kronecker power of $\mathbf{F}$. The codewords are now obtained by $\mathbf{x}=\mathbf{uG}_N$, where $\mathbf{u}$ contains $k$ information bits and $N-k$ frozen positions, for details we refer to \cite{ArikanPolar09}. This way, polar codes are inherently structured. 

\subsection{Design parameters of \ac{NND}}
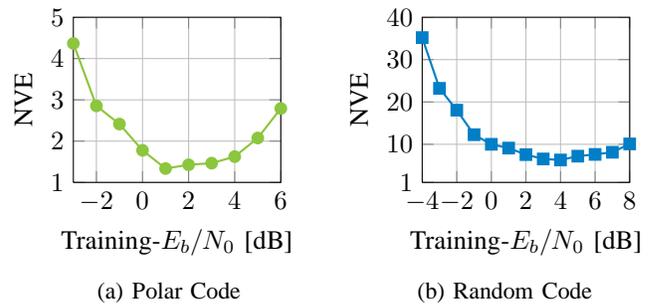
\begin{figure}
 \centering
 \begin{subfigure}{0.49\linewidth}
    \begin{tikzpicture}
\begin{axis}[
        width=\linewidth,
        height=0.16\textheight,
        xmajorgrids,
        yminorticks=true,
        ymajorgrids,
        yminorgrids,
        legend pos=north east,
        xlabel={Training-$E_b/N_0$ [dB]},
        ylabel={${\text{NVE}}$},
        xmin=-3,
        xmax=6,
        ymin=1,
        ymax=5,
        ytick={1,...,5}
    ]
    
    \addplot[color=apfelgruen, mark=*, thick] table {data/plot_trainingSNR_polar.data};
    \end{axis}

\end{tikzpicture}
    \caption{Polar Code}
    \label{fig:trainingSNR:polar}
 \end{subfigure} 
 \begin{subfigure}{0.49\linewidth}
    \begin{tikzpicture}
\begin{axis}[
        width=\linewidth,
        height=0.16\textheight,
        xmajorgrids,
        yminorticks=true,
        ymajorgrids,
        yminorgrids,
        legend pos=north east,
        xlabel={Training-$E_b/N_0$ [dB]},
        ylabel={${\text{NVE}}$},
        xmin=-4,
        xmax=8,
        xtick={-4,-2,0,2,4,6,8},
        ymin=1,
        ymax=40,
        ytick={1,10,20,30,40}
    ]

    \addplot[color=mittelblau, mark=square*, thick] table {data/plot_trainingSNR_random.data};
    \end{axis}

\end{tikzpicture}
    \caption{Random Code}
    \label{fig:trainingSNR:random}
 \end{subfigure}
 \caption{$\text{NVE}$ versus training-$E_b/N_0$ for \unit[16]{bit}-length codes for a 128-64-32 \ac{NN} trained with $M_{\text{ep}} = 2^{16}$ training epochs.}
 \label{fig:training_SNR} 
 \vspace{-0.3cm}
\end{figure}

Our starting point is a \ac{NN} as described before (see Fig.~\ref{fig:nn_setup}). We introduce the notation 128-64-32 which describes the design of the \ac{NN} decoder employing three hidden layers with $128$, $64$, and $32$ nodes, respectively. However, there are other design parameters with a non-negligible performance impact:
\begin{enumerate}
 \item What is the best training \ac{SNR}?
 \item How many training samples are necessary?
 \item Is it easier to learn from \ac{LLR} channel output values rather than from the direct channel output?
 \item What is an appropriate loss function?
 \item How many layers and nodes should the \ac{NN} employ?
 \item Which type of regularization\footnote{Regularization is any method that trades-off a larger training error against a smaller validation error. An overview of such techniques is provided in \cite[Ch. 7]{goodfellow2016}. We do not use any regularization techniques in this work, but leave it as an interesting future investigation.} should be used?
\end{enumerate}
The area of research dealing with the optimization of these parameters is called \emph{hyperparameter optimization} \cite{bergstra2011}. In this work, we do not further consider this optimization and restrict ourselves to a fixed set of hyperparameters which we have found to achieve good results. Our focus is on the differences between random and structured codes.


Since the performance of \ac{NND} depends not only on the \ac{SNR} of the  validation data set (for which the \ac{BER} is computed) but also on the \ac{SNR} of the training data set\footnote{It would also be possible to have a training data set which contains a mix of different \ac{SNR} values, but we have not investigated this option here. Recently, the authors in \cite{george2017deep} observed that starting at a high training SNR and then gradually reducing the SNR works well.}, we
define below a new performance metric, the \emph{\ac{NVE}}. Denote by $\rho_\text{t}$ and $\rho_\text{v}$ the \ac{SNR} (measured as $E_b/N_0$) of the training and validation data sets, respectively, and let $\text{BER}_\text{NND}(\rho_\text{t},\rho_\text{v})$ be the \ac{BER} achived by a \ac{NN} trained at $\rho_\text{t}$ on data with $\rho_\text{v}$. Similarly, let 
$\text{BER}_\text{MAP}(\rho_\text{v})$ be the \ac{BER} of \ac{MAP} decoding at \ac{SNR} $\rho_\text{v}$. For a set of $S$ different validation data sets with \acp{SNR} $\rho_{\text{v},1},\dots,\rho_{\text{v},S}$, the \ac{NVE} is defined as
\begin{align}
\text{NVE}(\rho_\text{t}) =  \frac1S \sum_{s=1}^S \frac{\text{BER}_\text{NND}(\rho_\text{t},\rho_{\text{v},s})}{\text{BER}_{\text{MAP}}(\rho_{\text{v},s})}.
\end{align}
The \ac{NVE} measures how good a \ac{NND}, trained at a particular \ac{SNR}, is compared to \ac{MAP} decoding over a range of different \acp{SNR}.
Obviously, for $\text{\ac{NVE}} = 1$, the \ac{NN} achieves \ac{MAP} performance, but is generally greater. In the sequel, we compute the \ac{NVE} over $S=20$ different 
\ac{SNR} points from \unit[0]{dB} to \unit[5]{dB} with a validation set size of 20000 examples for each \ac{SNR}.

We train our \ac{NN} decoder in so-called ``epochs''. In each epoch, the gradient of the loss function is calculated over the entire training set $\mathcal{X}$ using \emph{Adam}', a method for stochastic gradient descent optimization \cite{kingma2014}. Since the noise layer in our architecture generates a new noise realization each time it is used, the \ac{NN} decoder will never see the same input twice. For this reason, although the training set has a limited size of $2^k$ codewords, we can train on an essentially unlimited training set by simply increasing the number of epochs $M_{\text{ep}}$. However, this makes it impossible to distinguish whether the \ac{NN} is improved by a larger amount of training samples or more optimization iterations.

Starting with a \ac{NN} decoder architecture of 128-64-32 and $M_{\text{ep}}=2^{22}$ learning epochs, we train the \ac{NN} with datasets of different training \acp{SNR} and evaluate the resulting \ac{NVE}. The result is shown in Fig. \ref{fig:training_SNR}, from which it can be seen that there is an ``optimal'' training $E_b/N_0$. An explanation for the occurrence of an optimum can be explained by the two cases:
\begin{enumerate}
\item $E_b/N_0\to\infty$; train without noise, the \ac{NN} is not trained to handle noise.
\item $E_b/N_0\to0$; train only with noise, the \ac{NN} can not learn the code structure.
\end{enumerate}
This clearly indicates an optimum somewhere in between these two cases. From now on, a training $E_b/N_0$ of \unit[1]{dB} and \unit[4]{dB} is chosen for polar and random codes, respectively.

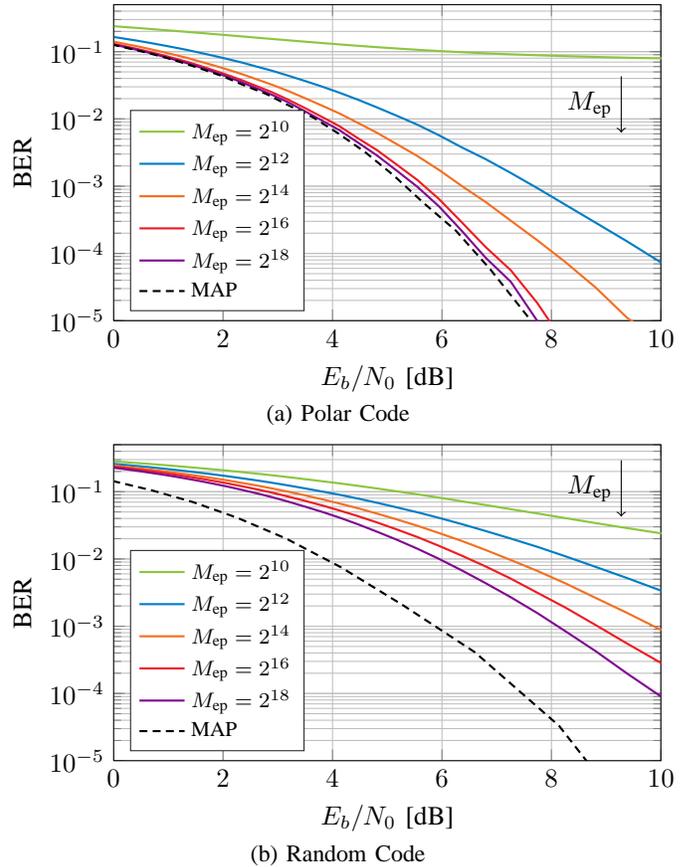
\begin{figure}[t]
 \centering
 \begin{subfigure}{\linewidth}
    \begin{tikzpicture}
\begin{axis}[
        width=\linewidth,
	height=0.245\textheight,
        xmajorgrids,
        yminorticks=true,
        ymajorgrids,
        yminorgrids,
        legend pos=south west,
        legend entries={$M_{\text{ep}} =2^{10}$, $M_{\text{ep}} =2^{12}$, $M_{\text{ep}} =2^{14}$, $M_{\text{ep}} =2^{16}$, $M_{\text{ep}} =2^{18}$, MAP},
        legend style={legend cell align=left,align=left,draw=white!15!black, font=\footnotesize},
        xlabel={$E_b/N_0$ [dB]},
        ylabel={BER},
        ymode=log,
        mark size=1pt,
        xmin=0,
        xmax=10,
        ymin=0.00001,
        ymax=0.5
    ]
    \addlegendimage{no markers, color=apfelgruen, thick}
    \addlegendimage{no markers, color=mittelblau, thick}
    \addlegendimage{no markers, color=orange, thick}
    \addlegendimage{no markers, color=rot, thick}
    \addlegendimage{no markers, color=lila, thick}
    \addlegendimage{color=black, densely dashed, thick}
    
    \addplot[color=apfelgruen, thick] table {data/plot_training_ber_polar_Eb_10.data};
    \addplot[color=mittelblau, thick] table {data/plot_training_ber_polar_Eb_12.data};
    \addplot[color=orange, thick] table {data/plot_training_ber_polar_Eb_14.data};
    \addplot[color=rot, thick] table {data/plot_training_ber_polar_Eb_16.data};
    \addplot[color=lila, thick] table {data/plot_training_ber_polar_Eb_18.data};
    
    \addplot[color=black, densely dashed, thick] table {data/bit_error_rate_Eb_map_polar.data};

\end{axis}
\draw[->] (6.75,3.25) -- node[left,midway] {$M_{\text{ep}}$} (6.75,2.5);
\end{tikzpicture}
    \vspace{-0.6cm}
    \caption{Polar Code}
    \vspace{0.2cm}
    \label{fig:training-size:polar}
 \end{subfigure} 
 \begin{subfigure}{\linewidth}
    \begin{tikzpicture}
\begin{axis}[
        width=\linewidth,
	height=0.245\textheight,
        xmajorgrids,
        yminorticks=true,
        ymajorgrids,
        yminorgrids,
        legend pos=south west,
        legend entries={$M_{\text{ep}} =2^{10}$, $M_{\text{ep}} =2^{12}$, $M_{\text{ep}} =2^{14}$, $M_{\text{ep}}=2^{16}$, $M_{\text{ep}} =2^{18}$, MAP},
        legend style={legend cell align=left,align=left,draw=white!15!black, font=\footnotesize},
        xlabel={$E_b/N_0$ [dB]},
        ylabel={BER},
        ymode=log,
        mark size=1pt,
        xmin=0,
        xmax=10,
        ymin=0.00001,
        ymax=0.5
    ]
    \addlegendimage{no markers, color=apfelgruen, thick}
    \addlegendimage{no markers, color=mittelblau, thick}
    \addlegendimage{no markers, color=orange, thick}
    \addlegendimage{no markers, color=rot, thick}
    \addlegendimage{no markers, color=lila, thick}
    \addlegendimage{color=black, densely dashed, thick}
    
    \addplot[color=apfelgruen, thick] table {data/plot_training_ber_random_Eb_10.data};
    \addplot[color=mittelblau, thick] table {data/plot_training_ber_random_Eb_12.data};
    \addplot[color=orange, thick] table {data/plot_training_ber_random_Eb_14.data};
    \addplot[color=rot, thick] table {data/plot_training_ber_random_Eb_16.data};
    \addplot[color=lila, thick] table {data/plot_training_ber_random_Eb_18.data};
    
    \addplot[color=black, densely dashed, thick] table {data/bit_error_rate_Eb_map_random.data};

\end{axis}
\draw[->] (6.75,4.0) -- node[left,midway] {$M_{\text{ep}}$} (6.75,3.25);
\end{tikzpicture}
    \vspace{-0.6cm}
    \caption{Random Code}
    \label{fig:training-size:random}
 \end{subfigure}
 \caption{Influence of the number of epochs $M_{\text{ep}}$ on the \ac{BER} of a 128-64-32 \ac{NN} for \unit[16]{bit}-length codes with code rate $r=0.5$.}
 \label{fig:training-size} 
 \vspace{-0.3cm}
\end{figure}

Fig.~\ref{fig:training-size} shows the \ac{BER} achieved by a  very small \ac{NN} of dimensions 128-64-32 as a function of the number of training epochs ranging from $M_{\text{ep}} = 2^{10}, \ldots, 2^{18}$. For \ac{BER} simulations, we use 1 million codewords per \ac{SNR} point. For both code families, the larger the number of training epochs, the closer is the gap between \ac{MAP} and \ac{NND} performance. However, for polar codes, close to \ac{MAP} performance is already achieved for $M_{\text{ep}}=2^{18}$ epochs, while we may need a larger \ac{NN} or more training epochs for random codes.

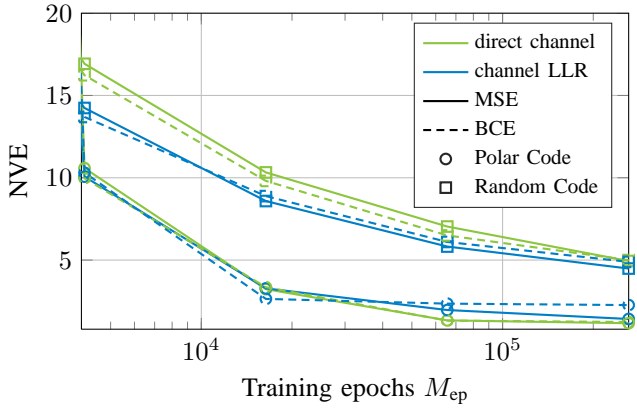
\begin{figure}[t]
 \centering
        \begin{tikzpicture}
        \begin{axis}[
            width=\linewidth,
            height=0.245\textheight,
            xmajorgrids,
            yminorticks=true,
            ymajorgrids,
            yminorgrids,
            legend pos=north east,
            legend entries={direct channel, channel LLR, MSE, BCE, Polar Code, Random Code},
            legend style={legend cell align=left,align=left,draw=white!15!black, font=\footnotesize},
            xlabel={Training epochs $M_{\text{ep}}$},
            ylabel={${\text{NVE}}$},
            xmode=log,
            xmin=4000,
            xmax=262144,
            ymin=0.8,
            ymax=20,
        ]
	\addlegendimage{no markers, color=apfelgruen, thick, solid}
	\addlegendimage{no markers, color=mittelblau, thick, solid}
	\addlegendimage{no markers, color=black, thick, solid}
	\addlegendimage{no markers, color=black, thick, densely dashed}
	\addlegendimage{only marks, mark=o, thick}
        \addlegendimage{only marks, mark=square, thick}
        
        \addplot[color=apfelgruen, mark=o, thick, solid] table {data/plot_llrloss_learning_polar_mse_channel.data};
        \addplot[color=mittelblau, mark=o, thick, solid] table {data/plot_llrloss_learning_polar_mse_llr.data};
        \addplot[color=apfelgruen, mark=o, thick, densely dashed] table {data/plot_llrloss_learning_polar_bce_channel.data};
        \addplot[color=mittelblau, mark=o, thick, densely dashed] table {data/plot_llrloss_learning_polar_bce_llr.data};
        
        \addplot[color=apfelgruen, mark=square, thick, solid] table {data/plot_llrloss_learning_random_mse_channel.data};
        \addplot[color=mittelblau, mark=square, thick, solid] table {data/plot_llrloss_learning_random_mse_llr.data};
        \addplot[color=apfelgruen, mark=square, thick, densely dashed] table {data/plot_llrloss_learning_random_bce_channel.data};
        \addplot[color=mittelblau, mark=square, thick, densely dashed] table {data/plot_llrloss_learning_random_bce_llr.data};

        \end{axis}
        \end{tikzpicture}
 \caption{Learning curve for \unit[16]{bit}-length codes with code rate $r=0.5$ for a 128-64-32 \ac{NN}.}
 \label{fig:llrloss} 
  \vspace{-0.55cm}
\end{figure}

In Fig. \ref{fig:llrloss}, we illustrate the influence of direct channel values versus channel \ac{LLR} values as decoder input in combination with two loss functions, \ac{MSE} and \ac{BCE}. The \ac{NVE} for all combinations is plotted as a function of the number of training epochs. Such a curve is also called ``learning curve'' since it shows the process of learning. Although it is ususally recommended to normalize the \ac{NN} inputs to have zero mean and unit variance, we train the \ac{NN} without any normalization which seems to be sufficient for our setup. For a few training epochs, the \ac{LLR} input improves the learning process; however, this advantage disappears for a larger $M_{\text{ep}}$. The same holds for \ac{BCE} against \ac{MSE}. For polar codes with \ac{LLR} values and \ac{BCE} the learning appears not to converge for the applied number of epochs. In summary, for training the \ac{NN} with a large number of training epochs it does not matter if \ac{LLR} or channel values are used as inputs and which loss function is employed. Moreover, normalization is not required.

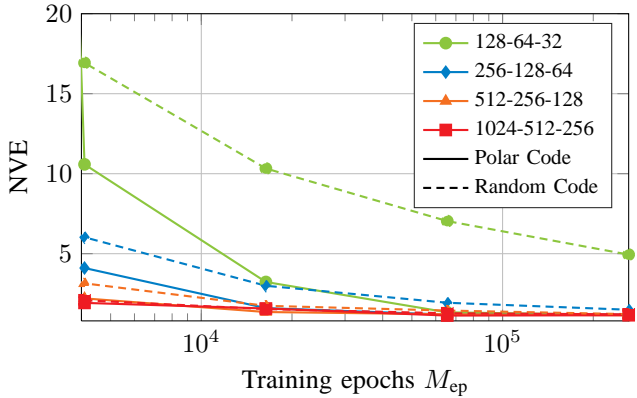
\begin{figure}[t]
 \centering
    \begin{tikzpicture}
\begin{axis}[
        width=\linewidth,
	height=0.24\textheight,
        xmajorgrids,
        yminorticks=true,
        ymajorgrids,
        yminorgrids,
        legend pos=north east,
        legend entries={128-64-32, 256-128-64, 512-256-128, 1024-512-256, Polar Code, Random Code},
        legend style={legend cell align=left,align=left,draw=white!15!black, font=\footnotesize},
        xlabel={Training epochs $M_{\text{ep}}$},
        ylabel={${\text{NVE}}$},
	xmode=log,
	mark size=2pt,
	xmin=4000,
        xmax=262144,
        ymin=0.8,
        ymax=20
    ]
    \addlegendimage{mark=*, color=apfelgruen, thick}
    \addlegendimage{mark=diamond*, color=mittelblau, thick}
    \addlegendimage{mark=triangle*, color=orange, thick}
    \addlegendimage{mark=square*, color=rot, thick}
    \addlegendimage{no markers,solid, color=black, mark=o, thick}
    \addlegendimage{no markers,densely dashed, color=black, mark=square, thick}
    
    \addplot[color=apfelgruen, mark=*, thick] table {data/plot_design_learning_polar_128.data};
    \addplot[color=mittelblau, mark=diamond*, thick] table {data/plot_design_learning_polar_256.data};
    \addplot[color=orange, mark=triangle*, thick] table {data/plot_design_learning_polar_512.data};
    \addplot[color=rot, mark=square*, thick] table {data/plot_design_learning_polar_1024.data};
    
    \addplot[color=apfelgruen, mark=*, thick,densely dashed] table {data/plot_design_learning_random_128.data};
    \addplot[color=mittelblau, mark=diamond*, thick,densely dashed] table {data/plot_design_learning_random_256.data};
    \addplot[color=orange, mark=triangle*, thick,densely dashed] table {data/plot_design_learning_random_512.data};
    \addplot[color=rot, mark=square*, thick,densely dashed] table {data/plot_design_learning_random_1024.data};

    \end{axis}
\end{tikzpicture}
 \caption{Learning curve for different \ac{NN} sizes for \unit[16]{bit}-length codes with code rate $r=0.5$.}
 \label{fig:design} 
 \vspace{-0.4cm}
\end{figure}

In order to answer the question how large the \ac{NN} should be, we trained \acp{NN} with different sizes and structures. From Fig.~\ref{fig:design}, we can conclude that, for both polar and random codes, it is possible to achieve \ac{MAP} performance. Moreover, and somewhat surprisingly, the larger the net, the less training epochs are necessary. In general, the larger the number of layers and neurons, the larger is the expressive power or \emph{capacity} of the \ac{NN} \cite{goodfellow2016}. Contrary to what is common in classic machine learning tasks, increasing the network size does not lead to overfitting since the network never sees the same input twice.

\subsection{Scalability}

\begin{figure}[t]
 \centering
    \begin{tikzpicture}
  \begin{axis}[
            width=\linewidth,
            height=0.21\textheight,
            xmajorgrids,
            yminorticks=true,
            ymajorgrids,
            yminorgrids,
            legend pos=north west,
            legend entries={$N=16$, $N=32$, $N=64$, Polar Code, Random Code},
            legend style={legend cell align=left,align=left,draw=white!15!black, font=\footnotesize},
            xlabel={\# information bits $k$},
            ylabel={${\text{NVE}}$},
            xmin=8,
            xmax=14,
            ymin=1,
            ymax=30
        ]
        \addlegendimage{color=apfelgruen, mark=*, thick}
	\addlegendimage{color=mittelblau, mark=diamond*, thick}
	\addlegendimage{color=orange, mark=triangle*, thick}
	\addlegendimage{no markers, color=black, thick}
	\addlegendimage{no markers, color=black, densely dashed, thick}
        
        \addplot[color=apfelgruen, mark=*, thick] table {data/plot_dimensionality_polarN=16.data};
        \addplot[color=mittelblau, mark=diamond*, thick] table {data/plot_dimensionality_polarN=32.data};
        \addplot[color=orange, mark=triangle*, thick] table {data/plot_dimensionality_polarN=64.data};
        
        \addplot[color=apfelgruen, mark=*, thick, densely dashed] table {data/plot_dimensionality_randomN=16.data};
        \addplot[color=mittelblau, mark=diamond*, thick, densely dashed] table {data/plot_dimensionality_randomN=32.data};
        \addplot[color=orange, mark=triangle*, thick, densely dashed] table {data/plot_dimensionality_randomN=64.data};
  \end{axis}
\end{tikzpicture}
 \caption{Scalability shown by \ac{NVE} for a 1024-512-256 \ac{NN} for \unit[16/32/64]{bit}-length codes with different code rates and ${M_{\text{ep}}=2^{16}}$ training epochs.}
 \label{fig:dimensionality} 
  \vspace{-0.4cm}
\end{figure}
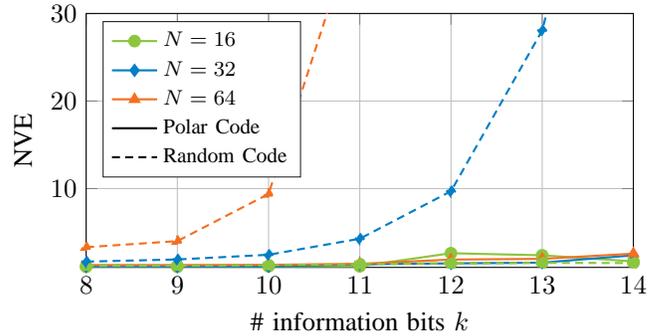

Up to now, we have only considered \unit[16]{bit}-length codes which are of little practical importance. Therefore, the scalability of the \ac{NN} decoder is investigated in Fig.~\ref{fig:dimensionality}. One can see that the length $N$ is not crucial to learn a code by deep learning techniques. What matters, however, is the number of information bits $k$ that determines the number of different classes ($2^k$) which the \ac{NN} has to distinguish. For this reason, the \ac{NVE} increases exponentially for larger values of $k$ for a \ac{NN} of fixed size and fixed number of training epochs. If a \ac{NN} decoder is supposed to scale, it must be able to generalize from a few training examples. In other words, rather than learning to classify $2^k$ different codewords, the \ac{NN} decoder should learn a decoding algorithm which provides the correct output for any possible codeword. In the next section, we investigate whether structure allows for some form of generalization.

\section{Capability of Generalization}
\begin{figure}[t]
 \centering
 \begin{subfigure}{\linewidth}
    \begin{tikzpicture}
\begin{axis}[
        width=\linewidth,
	height=0.25\textheight,
        xmajorgrids,
        yminorticks=true,
        ymajorgrids,
        yminorgrids,
        legend pos=south west,
        legend entries={$p=100\,\%$, $p=90\,\%$, $p=80\,\%$, $p=70\,\%$, MAP, $\overline{\mathcal{X}_{p}}$, $\mathcal{X}$ },
        legend style={legend cell align=left,align=left,draw=white!15!black, font=\footnotesize},
        xlabel={$E_b/N_0$ [dB]},
        ylabel={BLER},
        ymode=log,
        mark size=1.5pt,
        xmin=0,
        xmax=10,
        ymin=0.0001,
        ymax=1
    ]
    \addlegendimage{mark=*, color=apfelgruen, thick}
    \addlegendimage{mark=diamond*, color=mittelblau, thick}
    \addlegendimage{mark=triangle*, color=orange, thick}
    \addlegendimage{mark=square*, color=rot, thick}
    \addlegendimage{densely dotted, color=mittelgrau, thick}
    \addlegendimage{no markers, color=black, solid, thick}
    \addlegendimage{no markers, color=black, densely dashed, thick}
    
    \addplot[mark=*,color=apfelgruen, thick] table {data/bit_error_rate_coverage_polar_WER_Eb_1.data};
    
    \addplot[mark=diamond*,color=mittelblau, thick] table {data/bit_error_rate_coverage_polar_WER_Eb_0.9_inv.data};
    \addplot[mark=diamond*,color=mittelblau, densely dashed, thick] table {data/bit_error_rate_coverage_polar_WER_Eb_0.9_all.data};
    
    \addplot[mark=triangle*,color=orange, thick] table {data/bit_error_rate_coverage_polar_WER_Eb_0.8_inv.data};
    \addplot[mark=triangle*,color=orange, densely dashed, thick] table {data/bit_error_rate_coverage_polar_WER_Eb_0.8_all.data};
    
    \addplot[mark=square*,color=rot, thick] table {data/bit_error_rate_coverage_polar_WER_Eb_0.7_inv.data};
    \addplot[mark=square*,color=rot, densely dashed, thick] table {data/bit_error_rate_coverage_polar_WER_Eb_0.7_all.data};
    
    \addplot[color=mittelgrau, densely dotted, thick] table {data/bit_error_rate_coverage_polar_WER_Eb_map.data};

\end{axis}
\tikzstyle{label1} = [color=apfelgruen,font=\scriptsize]
\tikzstyle{label2} = [color=mittelblau,font=\scriptsize]
\tikzstyle{label3} = [color=orange,font=\scriptsize]
\tikzstyle{label4} = [color=rot,font=\scriptsize]

\node[label2] at (5.6,3) {90\,\%};
\node[label3] at (5.4,3.25) {80\,\%};
\node[label4] at (5.2,3.6) {70\,\%};
\node[label1] at (5,1.1) {100\,\%};

\node[label2] at (6.75,1.45) {90\,\%};
\node[label3] at (6.75,1.8) {80\,\%};
\node[label4] at (6.75,2.45) {70\,\%};
\end{tikzpicture}
    \vspace{-0.6cm}
    \caption{\unit[16]{bit}-length Polar Code ($r=0.5$)}
    \vspace{0.05cm}
    \label{fig:coverage:polar}
 \end{subfigure}
 \begin{subfigure}{\linewidth}
    \begin{tikzpicture}
\begin{axis}[
        width=\linewidth,
	height=0.25\textheight,
        xmajorgrids,
        yminorticks=true,
        ymajorgrids,
        yminorgrids,
        legend pos=south west,
        legend entries={$p=100\,\%$, $p=90\,\%$, $p=80\,\%$, $p=70\,\%$, MAP, $\overline{\mathcal{X}_{p}}$, $\mathcal{X}$ },
        legend style={legend cell align=left,align=left,draw=white!15!black, font=\footnotesize},
        xlabel={$E_b/N_0$ [dB]},
        ylabel={BLER},
        ymode=log,
        mark size=1.5pt,
        xmin=0,
        xmax=10,
        ymin=0.0001,
        ymax=1
    ]
    \addlegendimage{mark=*, color=apfelgruen, thick}
    \addlegendimage{mark=diamond*, color=mittelblau, thick}
    \addlegendimage{mark=triangle*, color=orange, thick}
    \addlegendimage{mark=square*, color=rot, thick}
    \addlegendimage{densely dotted, color=mittelgrau, thick}
    \addlegendimage{no markers, color=black, solid, thick}
    \addlegendimage{no markers, color=black, densely dashed, thick}
    
    \addplot[mark=*,color=apfelgruen, thick] table {data/bit_error_rate_coverage_random_WER_Eb_1.data};
    
    \addplot[mark=diamond*,color=mittelblau, thick] table {data/bit_error_rate_coverage_random_WER_Eb_0.9_inv.data};
    \addplot[mark=diamond*,color=mittelblau, densely dashed, thick] table {data/bit_error_rate_coverage_random_WER_Eb_0.9_all.data};
    
    \addplot[mark=triangle*,color=orange, thick,mark repeat={2}] table {data/bit_error_rate_coverage_random_WER_Eb_0.8_inv.data};
    \addplot[mark=triangle*,color=orange, densely dashed, thick] table {data/bit_error_rate_coverage_random_WER_Eb_0.8_all.data};
    
    \addplot[mark=square*,color=rot, thick,mark repeat={3}] table {data/bit_error_rate_coverage_random_WER_Eb_0.7_inv.data};
    \addplot[mark=square*,color=rot, densely dashed, thick] table {data/bit_error_rate_coverage_random_WER_Eb_0.7_all.data};
    
    \addplot[color=mittelgrau, densely dotted, thick] table {data/bit_error_rate_coverage_random_WER_Eb_map.data};

\end{axis}
\tikzstyle{label1} = [color=apfelgruen,font=\scriptsize]
\tikzstyle{label2} = [color=mittelblau,font=\scriptsize]
\tikzstyle{label3} = [color=orange,font=\scriptsize]
\tikzstyle{label4} = [color=rot,font=\scriptsize]

\node[label2] at (4.5,4.15) {90\,\%};
\node[label3] at (5.25,4.15) {80\,\%};
\node[label4] at (6,4.15) {70\,\%};

\node[label1] at (6.75,0.9) {100\,\%};
\node[label2] at (6.75,3.1) {90\,\%};
\node[label3] at (6.75,3.42) {80\,\%};
\node[label4] at (6.75,3.9) {70\,\%};
\end{tikzpicture}
    \vspace{-0.6cm}
    \caption{\unit[16]{bit}-length Random Code ($r=0.5$)}
    \label{fig:coverage:random}
 \end{subfigure}
 \caption{\ac{BLER} for a 128-64-32 \ac{NN} trained on $\mathcal{X}_p$ with $M_{\text{ep}}=2^{18}$ learning epochs. Solid and dashed lines show the performance on $\overline{\mathcal{X}_p}$ on $\mathcal{X}$, respectively.}
 \label{fig:coverage} 
  \vspace{-0.4cm}
\end{figure}
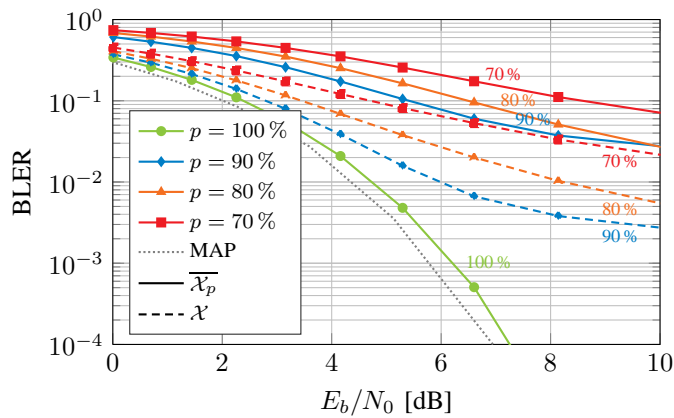
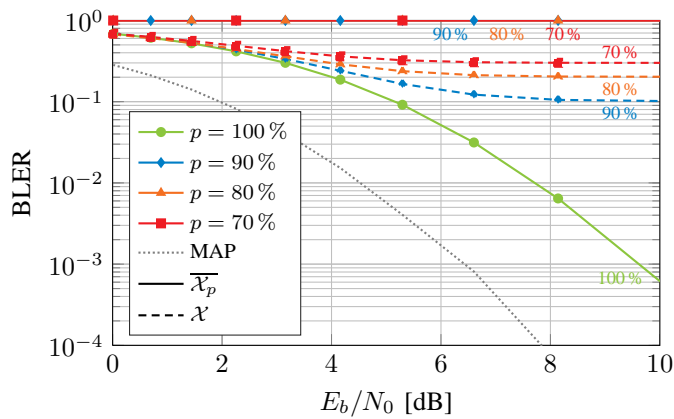

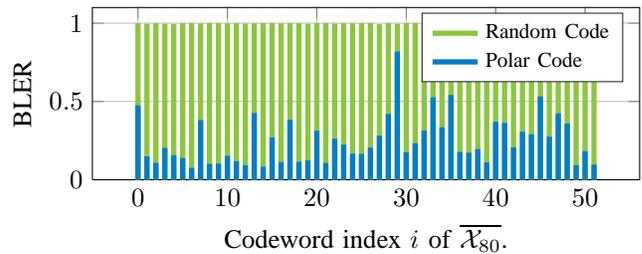
\begin{figure}
 \centering
 \begin{tikzpicture}
  \begin{axis}[
            width=\linewidth,
            height=0.164\textheight,
            xmajorgrids,
            yminorticks=true,
            ymajorgrids,
            yminorgrids,
            legend pos=north east,
            legend entries={Random Code, Polar Code},
            legend style={legend cell align=left,align=left,draw=white!15!black, font=\footnotesize},
            ylabel={BLER},
			ymin=0,
            xlabel={Codeword index $i$ of $\overline{\mathcal{X}_{80}}$.},
        ]
    \addlegendimage{color=apfelgruen, line width=2pt}
    \addlegendimage{color=mittelblau, line width=2pt}    
        
    \addplot[ycomb, color=apfelgruen, fill=apfelgruen, line width=2pt] table {data/histogram_coverage_random_single.data};
    
    \addplot[ycomb, color=mittelblau, fill=mittelblau, line width=2pt] table {data/histogram_coverage_polar_single.data};
    
  \end{axis}
\end{tikzpicture}
 \caption{Single-word \ac{BLER} for $\mathbf{x}_i \in \overline{\mathcal{X}_{80}}$ at $E_b/N_0 = \unit[4.16]{dB}$ and $M_{\text{ep}}=2^{18}$ learning epochs.}
 \label{fig:histogram} 
 \vspace{-0.4cm}
\end{figure}

As Fig.~\ref{fig:training_SNR}--\ref{fig:dimensionality} show, \acp{NND} for polar codes always perform better than random codes for a fixed \ac{NN} design and number of training epochs. This provides a first indication that structured codes, such as polar codes, are easier to learn than random codes. In order to confirm this hypothesis, we train the \ac{NN} based on a subset $\mathcal{X}_{p}$ which covers only p\,\% of the entire set of valid codewords. Then, the \ac{NN} decoder is evaluated with the set $\overline{\mathcal{X}_{p}}$ that covers the remaining $100 - p\,\%$ of $\mathcal{X}$. As a benchmark, we evaluate the \ac{NN} decoder also for the set of all codewords $\mathcal{X}$.  Instead of \ac{BER} as in Fig.~\ref{fig:training-size}, we now use the \ac{BLER} for evaluation (see Fig.~\ref{fig:coverage}). This way, we only consider whether an entire codeword is correctly detected or not, exluding side-effects of similarities between codewords which might lead to partially correct decoding. While for polar codes the \ac{NN} is able to decode codewords that were not seen during training, the \ac{NN} cannot decode any unseen codeword for random codes. Fig.~\ref{fig:histogram} emphasizes this observation by showing the single-word \ac{BLER} for the codewords $\mathbf{x}_i \in \overline{\mathcal{X}_{80}}$ which were not used for training. Obviously, the \ac{NN} fails for almost every unseen random codeword which is plausible. But for a structured code, such as a polar codes, the \ac{NN} is able to generalize even for unseen codewords. Unfortunately, the \ac{NN} architecture considered here is not able to achieve \ac{MAP} performance if it is not trained on the entire codebook. However, finding a network architecture that generalizes best is topic of our current investigations.

In summary, we can distinguish two forms of generalization. First, as described in Section \ref{sec:learn_to_decode}, the \ac{NN} can generalize from input channel values with a certain training \ac{SNR} to input channel values with arbitrary \ac{SNR}. Second, the \ac{NN} is able to generalize from a subset $\mathcal{X}_{p}$ of codewords to an unseen subset $\overline{\mathcal{X}_{p}}$. However, we observed that for larger \acp{NN} the capability of the second form of generalization vanishes. 

\section{Outlook and Conclusion}

For small block lengths, we achieved to decode random codes as well as polar codes with \ac{MAP} performance. But learning is limited through exponential complexity as the number of information bits in the codewords increases. The very surprising result is that the \ac{NN} is able to generalize for structured codes, which gives hope that decoding algorithms can be learned. State-of-the-art polar decoding currently suffers from high decoding complexity, a lack of possible parallelization and, thus, critical decoding latency. \ac{NND} inherently describes a highly parallelizable structure, enabling one-shot decoding. This renders deep learning-based decoding a promising alternative channel decoding approach as it avoids sequential algorithms.
Future investigations will be based on the exploration of regularization techniques as well as recurrent and memory-augmented neural networks, which are known to be Turing complete \cite{siegelmann1992computational} and have recently shown remarkable performance in algorithm learning. 

 \vspace{-0.0cm}
\bibliographystyle{IEEEtran}
\bibliography{references}
\end{document}